\documentstyle[twocolumn,psfig,longtable] {mn}

\begin{document}

\title[A high-resolution view of Cepheid atmospheres]
  {A photometric and spectroscopic study of the brightest northern
  cepheids. III. A high-resolution view of Cepheid atmospheres
  \thanks{Based on observations obtained at David Dunlap Observatory,
  Canada}}
\author[L.L. Kiss \& J. Vink\'o]
{L\'aszl\'o L. Kiss$^{1}$,
J\'ozsef Vink\'o$^2$\thanks{Hungarian E\"otv\"os Fellow, Department of
Astronomy, University of Toronto}\\
$^1$Department of Experimental Physics \& Astronomical Observatory,
JATE University, Szeged, D\'om t\'er 9, H-6720 Hungary,\\
E-mail: l.kiss@physx.u-szeged.hu\\
$^2$ Magyary Postdoctoral Fellow, Department of Optics \& Quantum Electronics, JATE University,\\
Research Group on Laser Physics of the Hungarian Academy of Sciences}

\maketitle

\begin{abstract}
We present new high-resolution spectroscopic observations
($\lambda / \Delta\lambda\approx$ 40000) of 18 bright northern
Cepheids carried out at David Dunlap Observatory, in 1997.
The measurements mainly extend those of presented in Paper I (Kiss 1998)
adding three more stars (AW~Per, SV~Vul, T~Mon).
The spectra were obtained in the yellow-red spectral region in
the interval of 5900 \AA$\ $and 6660 \AA, including strong lines of
sodium D and H$\alpha$. New radial velocities determined with
the cross-correlation technique and the bisector technique are presented.
The new data are compared with those recently published by several
groups. We found systematic differences between the spectroscopic
and CORAVEL-type measurements as large as 1--3 km~s$^{-1}$ in certain
phases.

We performed Baade-Wesselink analysis for CK~Cam discovered by the
Hipparcos satellite. The resulting radius is 31$\pm$1 R$_\odot$,
which is in very good agreement with recent period-radius relation
by Gieren et al. (1999). It is shown that the systematic
velocity differences do not affect the Baade-Wesselink radius
more than 1\% for CK~Cam.

Observational pieces of evidence of possible velocity gradient affecting
the individual line profiles are studied. The full-width at half
minimum (FWHM) of the metallic
lines, similarly to the velocity differences, shows a very characteristic
phase dependence, illustrating the effect of global compression in the
atmosphere. The smallest line widths always occur around the maximal
radius, while the largest FWHM is associated with the velocity reversal
before the minimal radius. Three first overtone pulsators
do not follow the general trend: the largest FWHM in SU~Cas and
SZ~Tau occurs after the smallest radius, during the expansion, while
in V1334~Cyg there are only barely visible FWHM-variations. The possibility
of a bright yellow companion of V1334~Cyg is briefly discussed.
The observed line profile asymmetries exceed the values predicted
with a simple projection effect by a factor of 2--3. This could be
associated with the velocity gradient, which is also supported by
the differences between individual line velocities of
different excitation potentials.

\end{abstract}

\section{Introduction}

Detailed understanding of Cepheid variables,
as strictly periodic pulsating stars with supersonic
atmospheric motions, requires spectroscopic
observations in a wide wavelength interval with high spectral resolution
and good phase coverage. The published high-resolution (at least
$\lambda / \Delta \lambda \sim 20000-30000$)
Cepheid spectroscopy addressed to the kinematic and dynamic phenomena
covers only a sample of few bright stars.
The recently presented extensive optical observations are as follows:
Wallerstein et al. (1992) -- W~Sgr, $\kappa$~Pav, S~Mus, S~Nor,
$\beta$~Dor, Y~Oph, U~Car;
Butler (1993) -- FF~Aql, $\delta$~Cep, $\eta$~Aql, X~Cyg;
Breitfellner \& Gillet (1993a,b,c) -- $\delta$~Cep, $\eta$~Aql, S~Sge, X~Cyg;
Sabbey et al. (1995) -- $\delta$~Cep, $\eta$~Aql, $\zeta$~Gem, X~Sgr;
Baldry et al. (1997) -- $\ell$~Car; Butler \& Bell (1997) -- $\delta$~Cep,
$\eta$~Aql, X~Cyg; Gillet et al. (1999) -- $\delta$~Cep. Obviously,
the overwhelming majority of the northern classical Cepheids --
except a few (3--4) stars -- suffers from the observational neglect in this
point of view.
This fact turned our attention to the topic and the first results
based on echelle spectroscopy were already published
in Kiss (1998).

The
problem of spectral line formation in a moving stellar
atmosphere was studied theoretically by Karp (1978). He showed
that the different Doppler-shifts observed in lines of different
strengths depend on the ionization balance variations as well
as on the velocity gradient. Butler (1993) presented
phase dependent velocity differences between lines of
different excitation and ionizations. He found
up to 5 km~s$^{-1}$ velocity residuals compared to high
excitation potential (EP) Fe I lines during the phase of
rapidly decreasing velocity. This important issue was
explored in much details by Sabbey et al. (1995).
They discussed the importance of line profile asymmetries
when measuring radial velocities. In an earlier paper
of Wallerstein et al. (1992) a comparison between different
measuring techniques was done with similar purposes.
One of the main conclusions in these papers is that
assigning a Doppler-shift to a specific portion of the
line profiles is a difficult and ambiguous task.
Recently, Vink\'o et al. (1999) studied the
limitations of precise radial velocity measurements
in Cepheid atmospheres and concluded that there is
a natural limit of a few km~s$^{-1}$, which could not be
overstepped. They also showed that CORAVEL-type radial
velocity (i.e. cross-correlation velocity smoothed over
the whole visible spectral region) curves may differ from
those obtained from a selected set of spectral lines.

Another question concerning the variable line profiles
is the atmospheric motions in classical Cepheids described
by the turbulence variations. Turbulence is considered
to be one of the most important line broadening mechanisms
besides rotation, although its physical meaning is still unclear.
CORAVEL-type measurements (Benz \& Mayor 1982, Bersier \& Burki 1996)
led to the determination of turbulence variations in certain
Cepheids, although these studies have neglected the
effect of possible velocity gradient. Bersier \& Burki (1996)
pointed out the similarity of acceleration and turbulence
curves with local maxima around the velocity reversal.
Breitfellner \& Gillet (1993a,b,c)
used a nonlinear, nonadiabatic model of pulsation
in order to determine an excess FWHM of the observed spectral
lines. They analyzed four stars with different periods
and found different turbulent velocity curves. Recently,
Gillet et al. (1999) presented FWHM variations for $\delta$~Cep
with excellent phase coverage, concluding that the strongest
peak in the turbulent velocity curve at $\phi$=0.85 is
associated with the global atmospheric compression. They also
suggested the possibility of turbulence amplification induced
by shock waves of moderate intensities.

The main aim of this paper is to present new spectroscopic
observations for a larger sample of Cepheid variables covering a
wider period range. The (above mentioned) earlier papers
dealt almost exclusively with fundamental pulsators,
while in the short-period domain there are a few
bright overtone Cepheids (e.g. SU~Cas, DT~Cyg, SZ~Tau). Bersier \& Burki
(1996) suggested the different turbulence behaviour of
classical and s-Cepheids as a possible mode-discriminator.
Therefore, beside the new radial velocity data, we examine
the fine spectral differences between fundamental and
overtone pulsators.

The paper is organised as follows. The observations are described in
Sect.\ 2. New radial velocities together with the earlier ones are
discussed in Sect.\ 3, Sect.\ 4 deals with the line profile analysis
(variations of the FWHM and the asymmetries along the pulsational
cycle). A summary is given in Sect.\ 5.

\section{Observations}

% Fig.1
\begin{figure*}
\begin{center}
\leavevmode
\psfig{figure=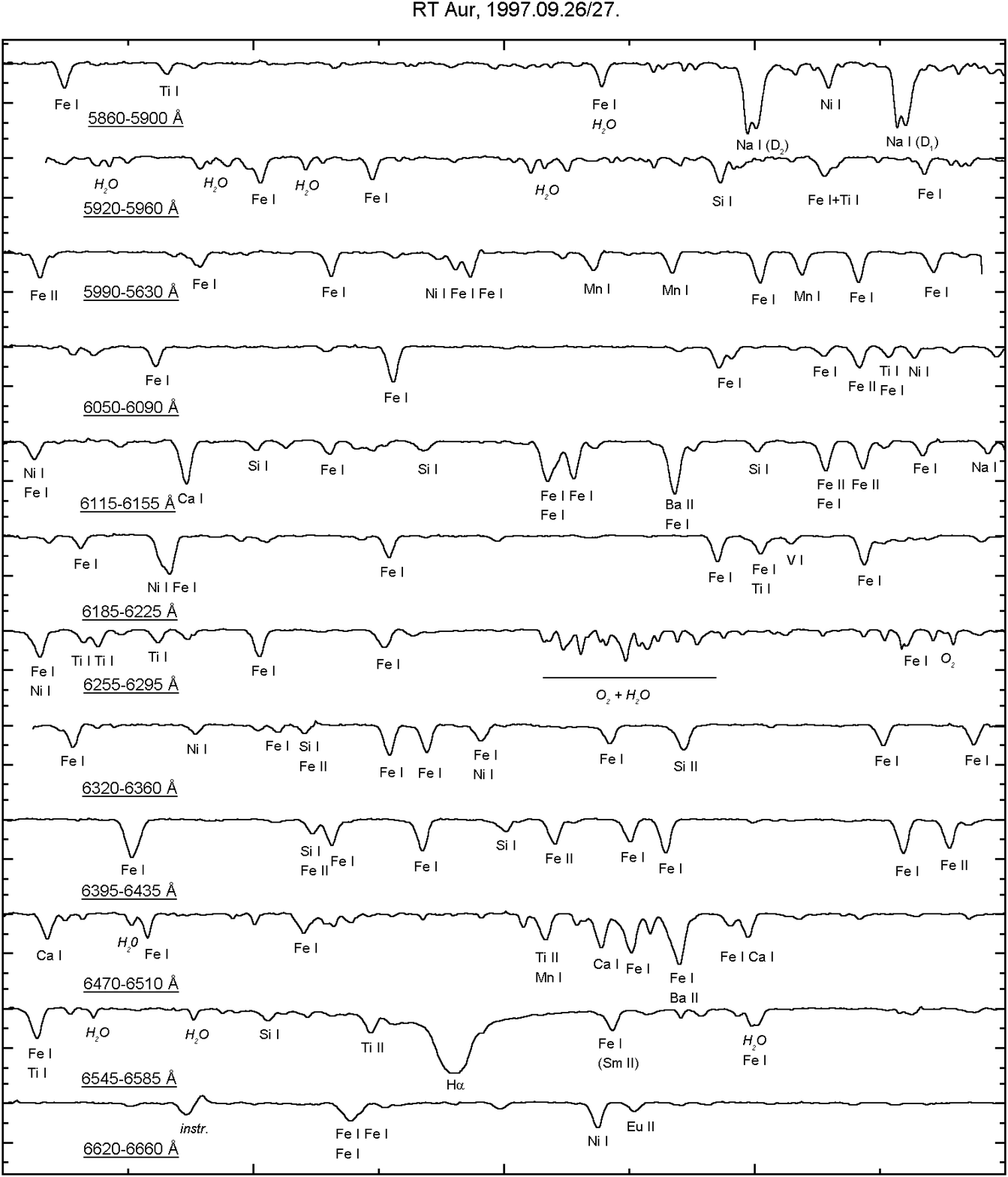,width=\textwidth}
\caption{A sample spectrum for RT~Aurigae.}
\end{center}
\end{figure*}

The high-resolution spectroscopic observations were carried out at
David Dunlap Observatory with the echelle spectrograph
attached to the 74-inch telescope in 1997 September. The
detector and the setup was the same as used by Kiss (1998), except
the cross-grating, which was replaced by the one with 600 lines/mm.
Therefore, we detected 12 orders with dispersion of 3.8 \AA\ mm$^{-1}$
giving a resolution of $\lambda/\Delta\lambda\approx40000$ at H$\alpha$.
All orders covered about 40 \AA\ between 5860 and 6660 \AA\ with
$\approx$25 \AA\ gaps between them. In this spectral region
we could identify more than 100 lines, the strong sodium D doublet,
H$\alpha$ and photospheric metal (mostly Fe I) lines. One order was
highly, another four were partially contaminated by atmospheric
telluric lines, which were monitored every night with observing
the rapidly rotating bright telluric standard HD~177724
(V=2.99 mag, spectral type A0V). As telluric lines change with airmass
and time, we did not apply telluric line corrections, only excluded the
affected spectral region from the further analysis. The exposure times were
between 5 and 40 min depending on the target brightness
and atmospheric conditions in order to reach a
signal-to-noise ratio of 50--250. The considerably high
cosmic ray contamination after 30 minutes did not allow the use
of longer exposures.

% Table 1.
\begin{table}
\begin{center}
\caption{The list of programme stars. Periods and epochs were
improved in five stars (typesetted in bold face)
using photometric data published in Paper I. The epoch and period
for T~Mon were taken from Evans et al. (1999). The remaining
ephemerides are from Szabados (1991) and Paper I (unchanged).}
\begin{tabular} {lllll}
\hline
Star & V$_{mean}$ & V$_{min}$ & Epoch & Period (d)\\
\hline
FF Aql & 5.18 & 5.68 & 50102.387 & 4.470936 \\
$\eta$ Aql & 3.48 & 4.39 & 50100.861 & 7.176726\\
RT Aur & 5.00 & 5.82 & 50101.159 & 3.728198\\
SU Cas & 5.70 & 6.18 & 50100.156 & 1.949325\\
{\bf $\delta$ Cep} & 3.48 & 4.37 & 50000.977 & 5.366316\\
{\bf X Cyg}  & 5.85 & 6.91 & 50007.597 & 16.38613 \\
DT Cyg & 5.57 & 5.96 & 50102.487 & 2.499086  \\
V1334 Cyg & 5.77 & 5.96 & 50102.549 & 3.332765 \\
{\bf $\zeta$ Gem} & 3.62 & 4.18 & 44232.443 & 10.1498 \\
S Sge  & 5.24 & 6.04 & 50105.348 & 8.382146  \\
SZ Tau & 6.33 & 6.75 & 50101.605 & 3.14873   \\
{\bf T Vul}  & 5.41 & 6.09 & 50101.410 & 4.4353  \\
SV Vul & 6.72 & 7.79 & 50104.50 & 45.0068    \\
AW Per  &  7.04 & 7.89 & 50103.361 & 6.463589 \\
CO Aur  & 7.46  & 8.08 & -- & --  \\
TU Cas  & 6.88  & 8.18 & -- & --  \\
CK Cam  & 7.23  & 7.81 & 50015.460 & 3.2942\\
{\bf T Mon} & 5.58  & 6.62 & 43784.615 & 27.024649\\
\hline
\end{tabular}
\end{center}
\end{table}

The list of observed stars is presented in Table\ 1. It contains
mostly the same stars as in Paper I. SV~Vul, AW~Per and
T~Mon was observed only in 1997, therefore, they were
not present in the spectroscopic programme of Paper I. We continued the
spectroscopic monitoring of two double-mode Cepheids, CO~Aur and TU~Cas.

The spectra were reduced with standard IRAF tasks including bias
removal, flat-fielding, cosmic ray elimination, echelle orders
extraction (with the task {\it doecslit}) and wavelength
calibration. For this calibration, we obtained two ThAr spectral
lamp exposures, immediately before and after every stellar
exposure. We applied careful linear interpolation between the
spectral lamp exposures, as slow wavelength shifts in the obtained spectra
were observed caused by the movement of telescope.
Each order was normalized to the continuum by fitting
cubic splines. The regions of strong lines (sodium D, H$\alpha$)
were omitted from the continuum fitting. A typical spectrum with
12 orders for RT~Aur is shown in Fig.\ 1, while the list of the
marked spectral lines is presented in Table\ 2. Note, that spectral
lines in long-period Cepheids (i.e. stars with lower effective
temperatures) are more numerous and stronger, thus Table\ 2 gives only a
rough overview of the obvious spectral features.

% Table 2.
\begin{table}
\begin{center}
\caption{List of identified strong (I$_{core}<0.9$) spectral lines. Lines
used for bisector velocity determination are typesetted in bold face.
Rest-wavelengths and excitation potentials were taken from Moore, Minnaert
\& Houtgast (1966).}
\begin{tabular} {llllll}
\hline
$\lambda$ &  ion   & EP & $\lambda$ & ion  & EP\\
          &     & (eV) &           &   & (eV)\\
\hline
5862.368 & Fe I         & 4.55&      6216.358 & V I          & 0.28\\                                  
5866.461 & Ti I         & 1.07&      {\bf 6219.287} & Fe I         & 2.20\\
5883.814 & Fe I         & 3.96&      6256.367 & Fe I         & 2.45\\                                     
5883.905 & {\it H$_2$0} & &                   & Ni I         & 1.68\\                                     
5889.973 & Na I (D$_2$) & 0.00 &     6258.110 & Ti I         & 1.44\\                                     
5892.883 & Ni I         & 1.99&      6258.713 & Ti I         & 1.46\\                                     
5895.940 & Na I (D$_1$) & 0.00&      {\bf 6261.106} & Ti I         & 1.43\\
5930.191 & Fe I         & 4.65&      {\bf 6265.141} & Fe I         & 2.18\\
5934.665 & Fe I         & 3.93&      6270.231 & Fe I         & 2.86\\                                     
5948.548 & Si I         & 5.08&      6290.974 & Fe I         & 4.73\\                                     
5952.726 & Fe I         & 3.98&      6322.694 & Fe I         & 2.59\\                                     
5953.170 & Ti I         & 1.89&      {\bf 6327.604} & Ni I         & 1.68\\
5956.706 & Fe I         & 0.86&      6330.852 & Fe I         & 4.73\\                                     
5991.378 & Fe II        & 3.15&      6331.953 & Si I         & 5.08\\                                     
5997.782 & Fe I         & 4.61&               & Fe II        & 6.22\\                                     
6003.022 & Fe I         & 3.88&      {\bf 6335.337} & Fe I         & 2.20\\
6007.317 & Ni I         & 1.68&      {\bf 6336.830} & Fe I         & 3.69\\
6007.968 & Fe I         & 4.65&      6338.880 & Fe I         & 4.79\\                                     
6008.566 & Fe I         & 3.88&      6339.118 & Ni I         & 4.15\\                                     
6013.497 & Mn I         & 3.07&      6344.155 & Fe I         & 2.43\\                                     
6016.647 & Mn I         & 3.07&      {\bf 6347.095} & Si II        & 8.12\\
6020.186 & Fe I         & 4.61&      {\bf 6355.035} & Fe I         & 2.84\\
6021.803 & Mn I         & 3.07&      {\bf 6358.687} & Fe I         & 0.86\\
6024.068 & Fe I         & 4.55&      6400.009 & Fe I         & 3.60\\                                     
6027.059 & Fe I         & 4.07&      6407.291 & Si I         & 5.87\\                                     
6056.013 & Fe I         & 4.73&               & Fe II        & 3.89\\                                     
6065.494 & Fe I         & 2.61&      {\bf 6408.026} & Fe I         & 3.69\\
6078.499 & Fe I         & 4.79&      {\bf 6411.658} & Fe I         & 3.65\\
6082.718 & Fe I         & 2.22&      6414.987 & Si I         & 5.87\\                                     
6084.105 & Fe II        & 3.20&      {\bf 6416.928} & Fe II        & 3.89\\
6085.257 & Ti I         & 1.05&      {\bf 6419.956} & Fe I         & 4.73\\
         & Fe I         & 2.76&      {\bf 6421.360} & Fe I         & 2.28\\
6086.288 & Ni I         & 4.26&      {\bf 6430.856} & Fe I         & 2.18\\
6116.198 & Ni I         & 4.09&      6432.683 & Fe II        & 2.89\\                                     
6116.246 & Fe I         & 4.26&      6471.668 & Ca I         & 2.52\\                                     
6122.226 & Ca I         & 1.89&      6475.632 & Fe I         & 2.56\\                                     
6125.026 & Si I         & 5.61&      6481.878 & Fe I         & 2.28\\                                     
6127.912 & Fe I         & 4.14&      6491.582 & Ti II        & 2.06\\                                     
         &              & 4.28&      6491.666 & Mn I         & 3.76\\                                     
6131.577 & Si I         & 5.61&      6493.788 & Ca I         & 2.52\\                                     
6131.858 & Si I         & 5.61&      6494.994 & Fe I         & 2.40\\                                     
6136.624 & Fe I         & 2.45&      6496.472 & Fe I         & 4.79\\                                     
6137.002 & Fe I         & 2.20&      6496.908 & Ba II        & 0.60\\                                     
6137.702 & Fe I         & 2.59&      6498.945 & Fe I         & 0.96\\                                     
6141.727 & Ba II        & 0.70&      6499.654 & Ca I         & 2.52\\                                     
         & Fe I         & 3.60&      6546.252 & Fe I         & 2.76\\                                     
6145.020 & Si I         & 5.61&               & Ti I         & 1.43\\                                     
6147.742 & Fe II        & 3.89&      6555.466 & Si I         & 5.98\\                                     
6147.834 & Fe I         & 4.07&      6559.576 & Ti II        & 2.05\\                                      
6149.249 & Fe II        & 3.89&      6562.808 & H$\alpha$    & 10.20\\                                    
6151.623 & Fe I         & 2.18&      6569.224 & Fe I         & 4.73\\                                     
6154.230 & Na I         & 2.10&               & (Sm II)      & 1.49\\                                     
{\bf 6187.995} & Fe I   & 3.94&      6575.037 & Fe I         & 2.59\\
6191.186 & Ni I         & 1.68&      6633.427 & Fe I         & 4.83\\                                     
6191.571 & Fe I         & 2.43&      6633.758 & Fe I         & 4.56\\                                     
{\bf 6200.321} & Fe I         & 2.61&      6634.123 & Fe I         & 4.79\\
{\bf 6213.437} & Fe I         & 2.22&      6643.638 & Ni I         & 1.68\\
6215.149 & Fe I         & 4.19&      6645.127 & Eu II        & 1.38\\
6215.22  & Ti I         & 2.69&               &              &     \\
\hline
\end{tabular}
\end{center}
\end{table}

\section{Radial velocities}

Radial velocities presented in Table\ 3 were determined with two
different methods. In order to compare the data with those of in Paper I,
we obtained a set of radial velocities by cross-correlating
two spectral regions
of the Cepheid and IAU standard star spectra with the IRAF task
{\it fxcor}. We chose HD~187691
(spectral type F8V, v$_{\rm rad}=+0.1\pm0.3$ km s$^{-1}$) as an
overall template star. The cross-correlated regions were those of
6188--6220 and 6405--6435 \AA, where 19 photospheric lines
were identified (see Table\ 2).

% Fig.2
\begin{figure}
\begin{center}
\leavevmode
\psfig{figure=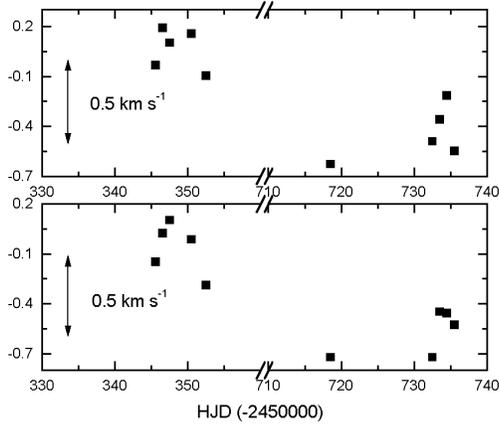,width=7cm}
\caption{Cross-correlation velocities from two spectral
orders for HD~187691 during the two observing seasons.
The systematic shift due to an instrumental problem (see text)
is 0.50 km s$^{-1}$.}
\end{center}
\end{figure}

%Table 3.
\begin{table*}
\begin{center}
\caption{The observed heliocentric radial velocities. $V_1$: cross-correlation
velocities; $V_2$: mean bisector velocities. The spectra of double-mode Cepheids
TU~Cas and CO~Aur did not allow determination of accurate single-line velocities
due to the increased observational scatter caused by the faintness of
the targets.}
\begin{tabular} {llrrllrrllrr}
\hline
Hel. J.D. & $\phi$ & $V_1$ & $V_2$ & Hel. J.D. & $\phi$ & $V_1$ & $V_2$ & Hel. J.D. & $\phi$ & $V_1$ & $V_2$\\
\hline
{\bf FF~Aql} & & &                          &      50344.659 & 0.059 & $-$33.89 & $-$34.73    &     50718.706 & 0.878 &  $-$4.65 &  $-$5.70       \\
50337.544 & 0.597 &  $-$8.31 & $-$8.61      &      50345.704 & 0.253 & $-$25.02 & $-$25.05   &     50731.737 & 0.788 &  $-$2.04 &  $-$3.46       \\
50345.540 & 0.385 & $-$16.54 & $-$16.63     &      50346.701 & 0.439 & $-$15.19 & $-$15.02    &     50732.653 & 0.063 &  $-$7.60 &  $-$9.13      \\
50346.532 & 0.607 &  $-$8.67 &  $-$8.82     &      50347.581 & 0.603 &  $-$6.22 &  $-$6.17    &     50733.681 & 0.372 &  $-$1.68 &  $-$3.47      \\
50347.525 & 0.829 & $-$14.40 & $-$14.84     &      50350.604 & 0.166 & $-$29.53 & $-$29.74    &     50734.652 & 0.663 &   1.59 &    $-$0.12      \\
50350.527 & 0.501 & $-$11.49 & $-$11.83     &      50352.544 & 0.528 & $-$10.26 & $-$10.25    &     50735.634 & 0.958 &  $-$7.40 &  $-$9.36      \\
50352.521 & 0.947 & $-$22.57 & $-$22.73     &      50355.739 & 0.123 & $-$31.73 & $-$32.23    &     50736.734 & 0.288 &  $-$3.14 &  $-$4.42      \\
50718.604 & 0.827 & $-$16.82 & $-$17.70     &      50356.779 & 0.317 & $-$21.77 & $-$21.97    &     50737.670 & 0.569 &   2.24 &    0.48         \\
50731.634 & 0.742 & $-$11.53 & $-$12.60     &      50718.721 & 0.764 &   2.01 &   1.28        &     50738.683 & 0.873 &  $-$5.45 &  $-$5.88      \\
50733.521 & 0.164 & $-$26.15 & $-$26.93     &      50731.754 & 0.192 & $-$28.79 & $-$29.01    &     {\bf $\zeta$~Gem} & & &                      \\
50734.512 & 0.386 & $-$18.27 & $-$19.09     &      50732.667 & 0.362 & $-$19.62 & $-$19.93    &     50337.854 & 0.530 &  21.49 &  21.35          \\
50735.551 & 0.618 & $-$10.83 & $-$11.27     &      50733.697 & 0.554 &  $-$9.27 &  $-$9.75    &     50345.875 & 0.320 &   7.74 &  7.82           \\
{\bf $\eta$~Aql} & & &                      &      50734.689 & 0.739 &   0.58 &  $-$0.24      &     50346.855 & 0.417 &  15.37 &  15.46          \\
50337.564 & 0.982 & $-$31.11 & $-$31.32     &      50735.504 & 0.891 & $-$11.78 & $-$12.26    &     50347.828 & 0.513 &  21.31 &  21.34          \\
50345.508 & 0.089 & $-$30.28 & $-$30.12     &      50735.707 & 0.929 & $-$25.07 & $-$25.30    &     50352.856 & 0.010 &  $-$1.59 &  $-$1.63      \\
50346.508 & 0.229 & $-$24.03 & $-$23.83     &      50737.683 & 0.297 & $-$23.37 & $-$23.93    &     50356.878 & 0.405 &  14.65 &   14.73         \\
50347.507 & 0.368 & $-$16.14 & $-$16.16     &      50738.660 & 0.478 & $-$13.76 &  $-$13.73    &     50718.878 & 0.070 &  $-$5.77 &  $-$6.17      \\
50350.503 & 0.785 &  10.16 &   9.32         &                &       &          &              &     50731.924 & 0.356 &  10.93 &   10.56        \\
50352.499 & 0.063 & $-$31.08 & $-$31.14     &      {\bf X~Cyg} & & &                           &     50732.913 & 0.453 &  18.31 &   17.81        \\
50355.677 & 0.506 & $-$13.78 & $-$13.80     &      50337.711 & 0.145 & $-$15.44 & $-$15.28     &     50733.869 & 0.547 &  21.95 &   21.19        \\
50356.620 & 0.637 &  $-$1.17 &  $-$1.31     &      50344.703 & 0.572 &  22.48 &   22.42        &     50737.864 & 0.941 &   0.20 &  $-$0.45      \\
50731.620 & 0.890 & $-$15.19 & $-$16.27     &      50345.656 & 0.630 &  28.30 &   27.68          &   {\bf S~Sge} & & &                          \\
50732.486 & 0.010 & $-$31.67 & $-$32.58     &      50346.661 & 0.692 &  33.35 &   32.16          &   50337.580 & 0.706 &  23.15 &  22.65         \\
50733.491 & 0.150 & $-$27.75 & $-$28.19     &      50347.595 & 0.749 &  35.87 &   34.71          &   50345.562 & 0.658 &  18.86 &  18.76         \\
50734.484 & 0.289 & $-$20.95 & $-$21.43     &      50350.616 & 0.933 &  12.75 &   12.47          &   50346.617 & 0.784 &  23.85 &  23.42         \\
50735.537 & 0.435 & $-$15.05 & $-$15.48     &      50352.601 & 0.054 & $-$20.51 & $-$20.13       &   50347.536 & 0.894 &   0.83 &  0.54          \\
50737.614 & 0.725 &   7.07 &   6.41         &      50355.727 & 0.245 &  $-$5.94 &  $-$5.85       &   50350.558 & 0.254 &  $-$4.19 &  $-$4.46     \\
{\bf RT~Aur} & & &                          &      50356.650 & 0.301 &  $-$0.80 &  $-$0.66       &   50352.554 & 0.492 &   1.41 &  1.63          \\
50337.829 & 0.481 &  24.99 &  25.20         &      50718.646 & 0.393 &   7.34 &   6.92           &   50355.709 & 0.869 &   6.46 &  6.35          \\
50345.863 & 0.635 &  33.27 &  33.28         &      50731.679 & 0.188 & $-$10.70 & $-$11.24       &   50356.665 & 0.983 & $-$12.59 & $-$12.81     \\
50346.824 & 0.893 &  22.84 &  22.87         &      50732.599 & 0.244 &  $-$5.24 &  $-$5.69       &   50718.624 & 0.165 & $-$36.28 & $-$36.72     \\
50347.795 & 0.154 &   6.78 &  6.69          &      50733.609 & 0.306 &   0.02 &  $-$0.32         &   50731.656 & 0.720 &  $-$3.42 &  $-$4.45     \\
50350.764 & 0.950 &   8.74 &  8.97          &      50734.553 & 0.364 &   5.26 &  4.78            &   50732.557 & 0.827 &  $-$7.84 &  $-$9.50     \\
50352.820 & 0.502 &  26.26 &  26.47         &      50735.521 & 0.422 &   9.65 &  9.31            &   50733.560 & 0.947 & $-$36.10 & $-$36.93     \\
50356.835 & 0.579 &  30.10 &  30.30         &      50737.649 & 0.552 &  20.69 &  20.18           &   50734.497 & 0.059 & $-$37.73 & $-$38.51     \\
50718.844 & 0.679 &  34.87 &  34.12         &      {\bf DT~Cyg} & & &                            &   50735.591 & 0.189 & $-$33.31 & $-$33.84     \\
50731.938 & 0.191 &   9.05 &  8.53          &      50337.628 & 0.091 &  $-$7.23 &  $-$7.34       &   50737.628 & 0.432 & $-$29.33 & $-$29.77     \\
50732.931 & 0.458 &  23.94 &  23.48         &      50337.752 & 0.140 &  $-$6.37 &  $-$6.27       &   {\bf SZ~Tau} & & &                          \\
50733.805 & 0.692 &  30.33 &  34.58         &      50344.746 & 0.939 &  $-$6.87 &  $-$7.07       &   50337.803 & 0.064 &  $-$7.03 &  $-$7.39     \\
50734.794 & 0.957 &   8.16 &  6.56          &      50345.615 & 0.287 &  $-$1.82 &  $-$1.85       &   50345.822 & 0.610 &   7.08 &   7.00         \\
50737.842 & 0.775 &  36.65 &  36.13         &      50345.759 & 0.344 &   0.09 &  0.03            &   50346.809 & 0.924 &  $-$2.43 &  $-$4.21 \\
50738.738 & 0.015 &   2.30 &  1.77          &      50346.604 & 0.683 &   5.21 &  5.17            &   50347.748 & 0.222 &  $-$5.46 &  $-$5.52 \\
{\bf SU~Cas} & & &                          &      50346.774 & 0.751 &   2.79 &  2.71            &   50350.745 & 0.174 &  $-$6.65 &  $-$7.49 \\
50337.667 & 0.843 &  $-$3.49 &  $-$3.41     &      50347.638 & 0.096 &  $-$7.17 &  $-$7.17       &   50352.800 & 0.827 &   5.77 &    5.53    \\
50345.581 & 0.903 &  $-$8.61 &  $-$8.60     &      50347.779 & 0.153 &  $-$6.09 &  $-$5.90       &   50356.813 & 0.101 &  $-$7.73 &  $-$9.03 \\
50345.739 & 0.984 & $-$13.43 & $-$14.03     &      50350.543 & 0.259 &  $-$2.82 &  $-$2.73       &   50718.826 & 0.072 &  $-$7.51 &  $-$8.22 \\
50346.574 & 0.412 &  $-$4.32 &  $-$4.34     &      50718.687 & 0.570 &   5.49 &  4.87            &   50731.879 & 0.217 &  $-$6.42 &  $-$6.45 \\
50346.756 & 0.506 &  $-$0.66 &  $-$0.61     &      50731.719 & 0.785 &   2.82 &  1.89            &   50732.766 & 0.499 &   2.20 &    2.23    \\
50347.652 & 0.966 & $-$12.44 & $-$12.61     &      50732.635 & 0.151 &  $-$5.47 &  $-$6.43       &   50733.789 & 0.824 &   7.82 &    7.00    \\
50347.878 & 0.082 & $-$15.39 & $-$15.91     &      50733.661 & 0.562 &   5.51 &    4.90          &   50734.722 & 0.120 &  $-$9.05 &  $-$9.08 \\
50350.685 & 0.521 &  $-$0.17 &  $-$0.28     &      50734.672 & 0.966 &  $-$6.35 &  $-$7.02       &   50737.746 & 0.081 &  $-$8.42 &  $-$9.05 \\
50352.844 & 0.629 &   1.75 &  1.75          &      50735.672 & 0.367 &   0.13 &    $-$0.44       &   50738.719 & 0.390 &  $-$1.67 &  $-$2.50 \\
50356.900 & 0.709 &   2.31 &  2.20          &      {\bf V1334~Cyg} & & &                         &   {\bf T~Vul} & & &                       \\
50718.743 & 0.334 &  $-$7.26 &  $-$7.64     &      50337.694 & 0.555 &  12.24 &  12.11           &   50337.651 & 0.262 &  $-$6.78 &  $-$6.80 \\
50731.785 & 0.024 & $-$15.57 & $-$16.83     &      50344.800 & 0.688 &  13.53 &  14.00           &   50344.720 & 0.856 &   6.45 &   5.99     \\
50732.684 & 0.486 &  $-$2.02 &  $-$2.48     &      50345.691 & 0.955 &   3.11 &  2.84            &   50345.676 & 0.071 & $-$16.77 & $-$16.83 \\
50733.717 & 0.016 & $-$15.45 & $-$16.49     &      50346.690 & 0.255 &   5.56 &  5.05            &   50346.677 & 0.297 &  $-$4.69 &  $-$4.72 \\
50734.582 & 0.459 &  $-$2.90 &  $-$3.49     &      50347.625 & 0.535 &  12.17 &  11.87           &   50347.611 & 0.507 &   6.77 &     6.66   \\
50737.718 & 0.068 & $-$15.02 & $-$16.36     &      50350.646 & 0.442 &  10.32 &  9.82            &   50350.632 & 0.189 & $-$10.77 & $-$10.90 \\
{\bf $\delta$~Cep} & & &                    &      50352.693 & 0.056 &   3.59 &  2.64            &   50352.676 & 0.649 &  13.06 &  13.09     \\
50337.735 & 0.768 &   2.61 &  2.3           &      50356.765 & 0.278 &   5.71 &  4.91            &   50356.682 & 0.552 &   8.69 &  8.75      \\
\hline
\end{tabular}
\end{center}
\end{table*}

\setcounter{table}{2}
%Table 3. cont.
\begin{table*}
\begin{center}
\caption{({\it cont.}) The observed heliocentric radial velocities.}
\begin{tabular} {llrrllrrllrr}
\hline
Hel. J.D. & $\phi$ & $V_1$ & $V_2$ & Hel. J.D. & $\phi$ & $V_1$ & $V_2$ & Hel. J.D. & $\phi$ & $V_1$ & $V_2$\\
\hline
{\bf T~Vul} cont. & & &                  &  50346.841 & -- &  20.30 &                &  50345.785 & 0.251 &  $-$8.24 &  $-$8.22  \\
50718.666 & 0.164 & $-$12.07 & $-$12.68  &  50347.812 & -- &   5.85 &                &  50346.714 & 0.533 &   7.86 &   7.52      \\
50731.701 & 0.103 & $-$15.97 & $-$16.65  &  50352.875 & -- &  14.94 &                &  50347.722 & 0.839 &   8.14 &   7.46      \\
50732.618 & 0.309 &  $-$4.09 &  $-$4.73  &  50356.861 & -- &   8.05 &                &  50350.702 & 0.744 &  13.73 &   13.24     \\
50733.639 & 0.540 &   8.55 &  7.93       &  50732.870 & -- &  $-$0.66 &              &  50352.743 & 0.363 &  $-$2.15 &  $-$2.34  \\
50734.631 & 0.763 &  15.68 &  14.76      &  {\bf TU~Cas} & & &                       &  50355.747 & 0.275 &  $-$6.74 &  $-$7.29  \\
50735.654 & 0.994 & $-$18.08 & $-$1865   &  50345.720 & -- & $-$33.37 &              &  50356.704 & 0.565 &   8.68 &    8.28     \\
50738.699 & 0.680 &  13.07 &  13.14      &  50346.554 & -- & $-$16.05 &              &  50718.799 & 0.466 &   3.94 &    3.28     \\
{\bf SV~Vul} & & &                       &  50346.741 & -- & $-$15.07 &              &  50731.806 & 0.414 &  $-$0.39 &  $-$0.65  \\
50732.577 & 0.954 &  $-$6.95 &  $-$9.62  &  50347.676 & -- & $-$27.53 &              &  50732.815 & 0.720 &  14.51 &    13.23    \\
50734.605 & 0.999 & $-$21.41 & $-$22.35  &  50347.863 & -- & $-$37.12 &              &  50733.853 & 0.035 & $-$17.58 & $-$18.18  \\
50735.612 & 0.021 & $-$23.70 & $-$24.07  &  50350.667 & -- & $-$11.83 &              &  50737.778 & 0.226 &  $-$9.31 &  $-$10.10 \\
{\bf AW~Per} & & &                       &  50352.769 & -- & $-$25.24 &              &  {\bf T~Mon} & & &                        \\
50718.771 & 0.212 & $-$14.01 & $-$14.66  &  50356.744 & -- & $-$22.04 &              &  50718.862 & 0.590 &  34.98 & 33.90       \\
50731.832 & 0.232 & $-$13.97 & $-$13.96  &  50356.928 & -- & $-$18.74 &              &  50731.910 & 0.073 &  $-$2.14 &  $-$2.48  \\
50732.789 & 0.381 &  $-$5.28 &  $-$5.73  &  50718.903 & -- & $-$17.27 &              &  50732.895 & 0.109 &  $-$0.65 &  $-$0.85  \\
50733.767 & 0.532 &   1.72 &   $-0.84$   &  50732.710 & -- & $-$37.94 &              &  50733.827 & 0.143 &   1.58 &  1.15       \\
50737.804 & 0.157 & $-$16.79 & $-$16.44  &  50733.742 & -- & $-$17.52 &              &  50734.770 & 0.178 &   4.01 &  3.42       \\
{\bf CO~Aur} & & &                       &  {\bf CK~Cam} & & &                       &  50737.888 & 0.294 &  12.64 &  12.08      \\
50345.842 & -- &   5.84 &                &  50337.776 & 0.820 &   9.80 &  9.54       &            &       &        &             \\
\hline
\end{tabular}
\end{center}
\end{table*}

The stability of radial velocity measurements was tested by the
standard star itself. We plotted the cross-correlation velocities
in Fig.\ 2, where a systematic shift of 0.50 km s$^{-1}$
is present between data obtained in 1996 and 1997. The necessary
shift for matching the data is the same for the different wavelength
regions. Furthermore, other non-variable stars ($\alpha$~Per,
HD~22484) showed similar shift. This turned out to be an instrumental effect,
probably the entrance slit of the spectrograph was not uniformly illuminated
by the spectral lamp. Therefore, a systematic asymmetry occured
in the emission line profiles of the ThAr lamp introducing a
systematic subpixel-shift in the wavelength calibration.
We determined the mean correction of 0.5 km~s$^{-1}$ and applied it
to the calculated radial velocities. In order
to guarantee the homogeneity of the data reduction procedure,
we have also re-reduced our spectra obtained in 1996.
The newly calculated radial velocities do not differ from values
in Kiss (1998) more than 0.1--0.14 km~s$^{-1}$, the mean deviation is
about $-$0.05 km~s$^{-1}$.
This is less by a factor of 3 than the internal accuracy of measurements
estimated to be about $\pm$0.15 km~s$^{-1}$.
All data are tabulated in Table\ 3.

Three stars (FF~Aql, V1334~Cyg and S~Sge) show remarkable
changes of the $\gamma$-velocity due to orbital motion in
binary sytems. The best example with good phase coverage is
V1334~Cyg, where V$_\gamma$ changed about 11 km~s$^{-1}$
in one year. This is almost a half of the full
orbital velocity amplitude,
which is 27.8 km~s$^{-1}$, according to Evans (1995).
The sytematic velocity changes in FF~Aql and S~Sge are
2.3 km~s$^{-1}$ and 27.7 km~s$^{-1}$, respectively. These
stars have well-determined orbits (see Evans 1995),
thus we did not try to re-calculate their orbital elements.

\subsection{Comparison of velocity curves}

% Fig.3
\begin{figure}
\begin{center}
\leavevmode
\psfig{figure=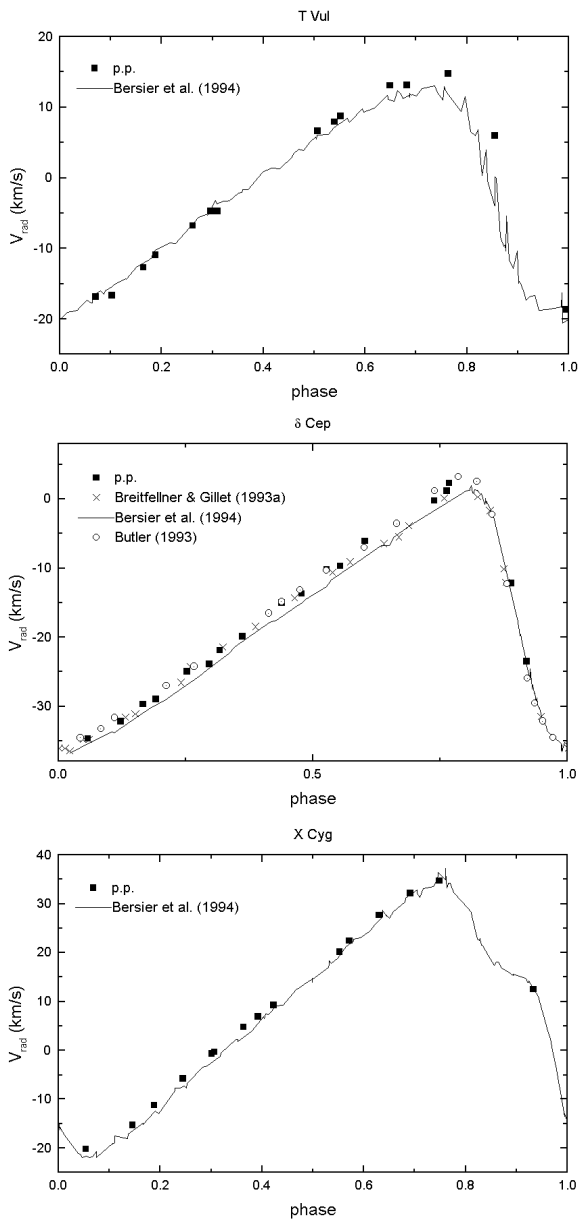,width=\linewidth}
\caption{Comparison of recently published radial velocities for
T~Vul, $\delta$~Cep and X~Cyg.}
\end{center}
\end{figure}

It is interesting to compare our new cross-correlation velocity
curves with those obtained by others with other instruments 
for the same stars, especially with the CORAVEL-type velocities
(Bersier et al. 1994), because those have been extensively
used in many studies of Cepheid variables. This comparison can
be seen in Fig.\ 3 for T~Vul, $\delta$~Cep and X~Cyg
where the continuous lines mean CORAVEL-velocities
and symbols denote others' data including ours. These three stars 
have well-covered velocity curves and they are used in order to 
represent shorter and longer period Cepheids. In the case of 
$\delta$~Cep data of Breitfellner \& Gillet (1993a) and Butler
(1993) were also plotted. The velocities of Breitfellner \& Gillet
are based on a few selected Fe I lines, while Butler's data have
been obtained with the very precise iodine-cell technique.
Note, that Butler (1993) reported that his data have no absolute
zero-point, and a 2--3 km~s$^{-1}$ shift is necessary to match his
$\delta$~Cep-velocities with those of others. We also added
a 3 km~s$^{-1}$ correction to these velocities and found perfect
agreement with our ccf-velocity curve as well as 
the velocities of Breitfellner \& Gillet (1993a). Contrary to 
these, there is a significant difference between all these
velocities and the CORAVEL data on the ascending branch of the
velocity curve of $\delta$~Cep.

An earlier version of this graph appeared in Paper I showing larger
differences between the datasets. Those larger shifts were partly
due to a slightly longer period used in Paper I for $\delta$~Cep.
In this paper we used an updated, shorter period (see Table 1)
that eliminated the larger dispersion of the phased velocities
on the ascending branch. However, a shift between the CORAVEL
and other types of velocities is still present. The same can be
observed in the case of X~Cyg and T~Vul, although for T~Vul the
largest differences occur during the velocity reversal.
As in the previous case, phase shifts due to slightly incorrect
epochs or periods have been eliminated by carefully taking into 
account the period variations of these stars, therefore all
the differences can be considered real. Moreover, Vink\'o et 
al. (1998) found similar deviations between
digital ccf- and CORAVEL-velocities of Type II Cepheids. 

All of these comparisons are consistent with each other and suggest
that for Cepheid variables the CORAVEL-type velocities systematically
differ from those obtained by direct spectral line measurements.
Possible reasons for the cause of this discrepancy are the well-known
effects that perturb the line profiles of pulsating variables, 
namely the velocity differences and/or the line asymmetries.
These effects are phase-dependent, so they may be capable of
explaining why the deviations of the CORAVEL-velocities get stronger
at certain phases. On the other hand, it is a bit surprising
that our digital cross correlation (ccf-) velocities 
(and also those of Vink\'o et al., 1998)
agree much better with line bisector- or iodine cell-velocities
than with CORAVEL-type ones. The digital cross-correlation
velocities should be (and probably they actually are) also
sensitive to the line profile disturbances mentioned above.
It is possible that another systematic effects are also present
in the CORAVEL-type measurements (perhaps associated with the
alignment of the mask or the usage of much more spectral features
in a much wider wavelength interval) that contribute to this 
disagreement. The velocity data obtained with CORAVEL have very
good phase coverage and also very good inner precision, therefore
they contain very relevant information on pulsating stars.
Therefore we conclude that the CORAVEL data should be compared
with other velocity curves obtained with other techniques before
using them for deriving astrophysical information that may be 
sensitive to a few km~s$^{-1}$ systematic differences.

\subsection{Line level effects and velocity differences}

% Fig.4
\begin{figure*}
\begin{center}
\leavevmode
\psfig{figure=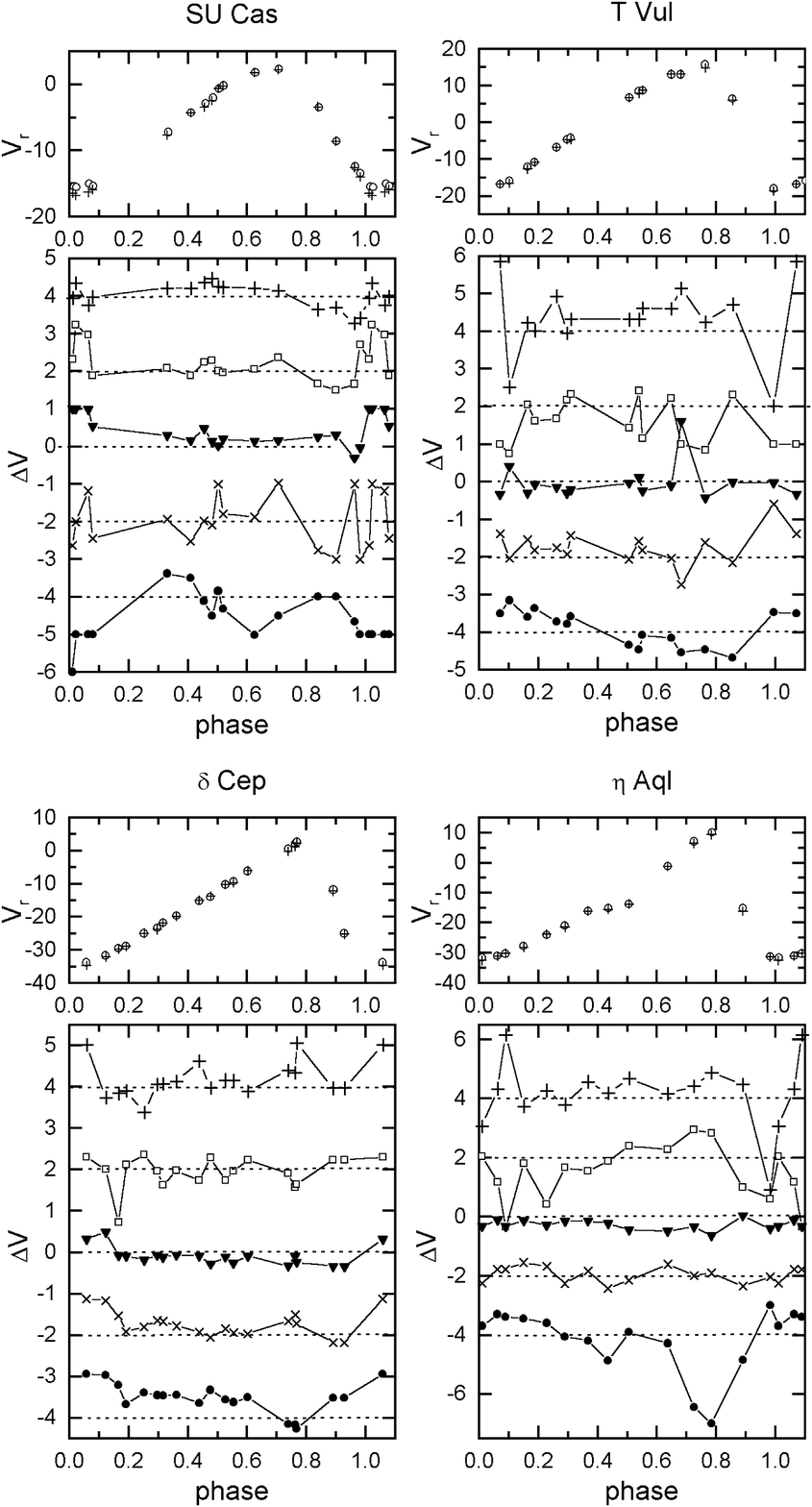,width=10cm}
\caption{Radial velocity data for SU~Cas, T~Vul, $\delta$~Cep and $\eta$~Aql.
Upper panels show the cross-correlation (open circles) and
mean bisector velocities (pluses). Differences between the individual
and mean bisector data for selected lines (pluses -- 6358.6879 \AA, Fe I 0.86
eV; open squares -- 6327.604 \AA, Ni I 1.68 eV; solid down triangles --
6411.658 \AA, Fe I 3.65 eV; crosses -- 6408.026 \AA, Fe I 3.69 eV; solid
circles -- 6347.095 \AA, Si II 8.12 eV) are plotted in the lower panels.
A radial velocity offset of 2 km~s$^{-1}$ was applied for clarity.}
\end{center}
\end{figure*}

% Fig.5
\begin{figure*}
\begin{center}
\leavevmode
\psfig{figure=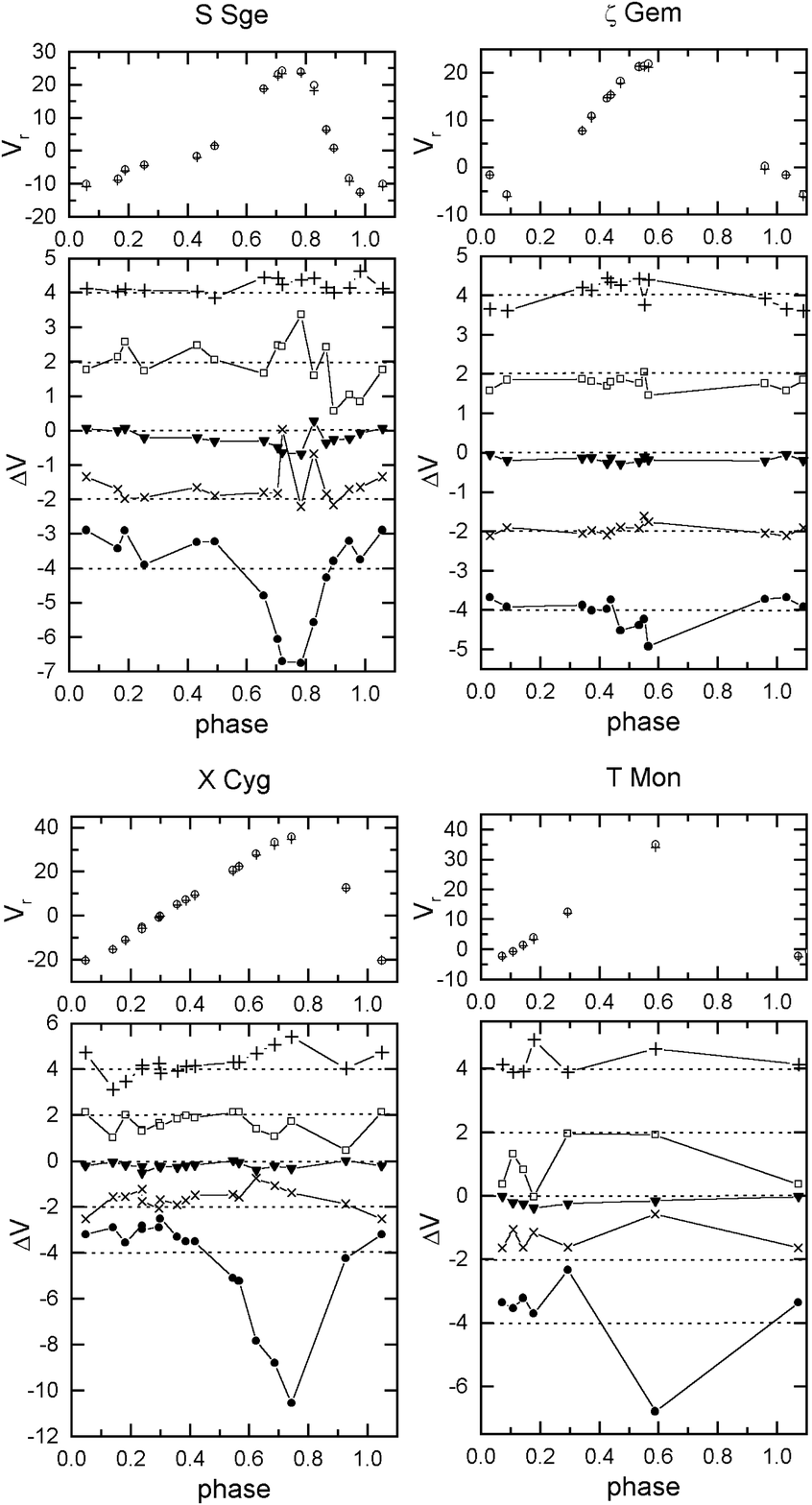,width=10cm}
\caption{Radial velocity data for S~Sge, $\zeta$~Gem, X~Cyg and T~Mon.
The symbols used are the same as in Fig.\ 4.}
\end{center}
\end{figure*}

Single-line radial velocities were also calculated for a set of
well-defined, unblended lines observed in both seasons. We used the
line bisector technique (see Wallerstein et al. 1992 for the definition)
to derive radial velocities measuring
Doppler-shifts of unblended Fe~I, Ni~I, Ti~I and Si~II lines in
the 6100--6450 \AA\ region. We chose the 0.7 bisector, because our
tests showed that it gives the most accurate velocities. The precision
of the individual line velocities were determined using artificial
digital spectra with random noise co-added. The precision turned
out to be $\pm500 {\rm m~s^{-1}}$ and $\pm150 {\rm m~s^{-1}}$ for
spectra of S/N=50 and S/N=200, respectively (Vink\'o et al. 1999).
Mean radial velocities were
calculated and plotted in Figs.\ 4--5 with the corresponding
ccf data ($V_1$) for eight stars covering a wide
period range (1.9--27 days). Radial
velocity differences between lines of different excitation
potentials (EP) and the mean metallic velocities (in sense of
V$_{\rm line}$ minus V$_{\rm mean}$) are also plotted in the
subpanels of Figs.\ 4--5. The referee suggested to also include
H$\alpha$ velocities, however, the H$\alpha$ profile is strongly
affected by various atmospheric phenomena and, consequently, it is
quite misleading to associate one velocity to the distorted line profile
(see, e.g. Wallerstein et al. 1992, Vink\'o et al. 1998). A detailed
analysis of H$\alpha$-observations, especially concerning the H$\alpha$
emission in long-period Cepheids, will be presented in a forthcoming
paper (Vink\'o \& Kiss, in preparation). Here we want to focus on the
velocity behaviour of metallic lines.

A few important conclusions can be drawn based on Figs.\ 4--5. First,
the {\it average} 0.7 bisector velocities are in
good agreement with the ccf data at level of 0.5--1 km~s$^{-1}$. This
is an expected result, since the cross-correlation technique results in
a mean Doppler-shift for lines in the selected wavelength region.
Second, there are clear examples of line level effect, especially
for low-EP ($<2$ eV) and high-EP ($>6$ eV) lines. Our results are very
similar to those of presented by Butler (1993) for 4 stars. The
Si~II 6347.1 \AA\ line is a very good indicator of the strength of
level effects. Its deviations, however, cannot be explained by
kinematic velocity gradients (Butler et al. 1996), because the
$\gamma$-velocity of the Si~II-velocities is quite different from
the $\gamma$-velocity of the average velocity curve.

It is visible in the lower panels of Figs.\ 4--5 that the $\gamma$-
velocities of some other lines also differs from the $\gamma$-
velocity of the average radial velocity curve. This was also
reported by Butler (1993), and it gives further support to
the conclusion of Sabbey et al. (1995) that path conservation
may not be valid for the integral of the observed velocity 
curves of Cepheids over the whole pulsational cycle.

The level effects get stronger with increasing period starting from
1--2 km~s$^{-1}$ for $P<7$ days up to 6--8 km~s$^{-1}$ for $P>7$
days. Note that the $\pm$2 km~s$^{-1}$ scattering in the velocity
differences of SU~Cas and T~Vul at later phases is due to
observational uncertainties rather than real physical effects. 
In the spectra of these stars the lowest and highest EP lines
become very weak at these phases which degrades the precision of
the velocity measurement.
In the case of the other stars the observed velocity differences
can be considered significant, especially for
the longer period ones. Unfortunately, the data of T~Mon have
very bad phase coverage, but the observed tendencies suggest 
similar behaviour to X~Cyg, namely the low-EP lines have positive
velocity residuals while the high-EP lines have large negative
velocity residuals in the 0.6--1.0 phase interval.

We tried to give an order of magnitude estimate of the dependence
of the velocity {\it gradients} on the stellar parameters from 
the observed velocity differences. This requires the knowledge
of the geometric distance between the line forming regions of
low- and high-EP lines, which is difficult to determine in a
dynamic atmosphere. As a very crude first approximation, 
we assumed that the lowest and highest E.P. lines form at the 
top and at the bottom of the photosphere, and that the width
of the photosphere is inversely proportional to the gravity,
thus $\Delta x_p \approx \alpha g^{-1}$ where $\alpha$ is roughly
constant (Gray, 1992). The absolute value of the velocity
gradient is then $\nabla v \approx \Delta v / \Delta x_p \approx
p \Delta v_r g / \alpha$ where $p$ is the conversion factor of radial
to pulsational velocities. We chose $\Delta v_r = v_r$(FeI 0.86 eV)
$ - v_r$(SiII 8.12 eV) and a constant $p$-factor of $p =$1.36.
Therefore we obtain
$\log \nabla v = \log \Delta v_r + \log g + {\rm constant}$.
It is visible that the velocity gradient has weaker
dependence on the pulsational period or the radius of the
Cepheid, because while $\Delta v_r$ increases with increasing period
(or radius via the period-radius relation), the gravity decreases 
at the same time. Thus, the velocity gradients are expected to
be roughly the same for short- and long-period Cepheids, provided
the simple scaling with gravity is indeed valid. Using the gravity
values presented in Paper II (Kiss \& Szatm\'ary 1998) it can be seen that the
velocity gradients calculated in this way have the same order 
of magnitude in the case of $\delta$ Cep and X Cyg. The absolute
value can be estimated assuming further that the thickness of
the Cepheid photospheres can be approximated with the scaling
of the width of the solar photosphere (about 500 km) to 
$\log g = 1 - 2$, resulting in a value of $\nabla v = 10^{-5}$
s$^{-1}$. However, in the case of strong shock waves forming in the
photosphere, the velocity gradient can be much larger than this,
at least temporarily.

\subsection{Baade-Wesselink radius of CK~Cam}

The Baade-Wesselink analysis is a widely used tool for radius determination
of pulsating variables. Here we do not want to discuss the recent
theoretical and observational efforts which have been done to
clarify the problems of this method, since it is
beyond the scope of present paper. We recall e.g. Gautschy (1987) for a
general review.
On the other hand, recent surface-brightness techniques (e.g.
Gieren et al. 1997) are based on different infrared colours,
which enables a reliable separation of temperature and radius
variation.

This subsection is addressed to CK~Cam, which was
discovered by the Hipparcos satellite (Makarov et al. 1994).
Because of the relatively new discovery of this star, there
are only few available photometric observations. Thus,
the sophisticated Baade-Wesselink implemantations or
surface-brightness methods cannot be used. That is why
we performed a simpler analysis following the original
assumptions of Wesselink (1946). This approach was used
recently by Balog et al. (1997) for Type II Cepheids, and
it is based on BV photometry and a full radial velocity curve.

% Fig.6
\begin{figure}
\begin{center}
\leavevmode
\psfig{figure=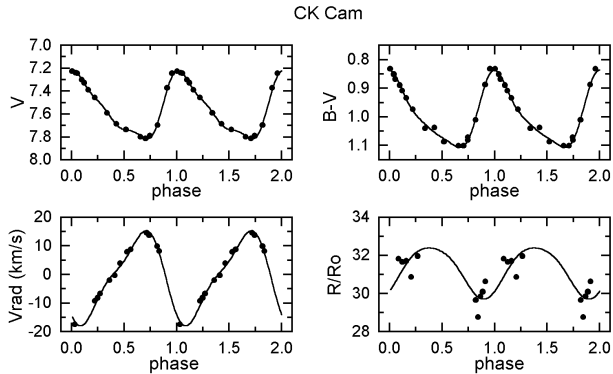,width=\linewidth}
\caption{Baade-Wesselink analysis of CK~Cam. }
\end{center}
\end{figure}

We have taken the published BV photometry of Berdnikov
et al. (1996). In order to reduce the numerical
uncertainties, we have fitted low-order (3--5) Fourier-polynomials
to the light, colour and radial velocity curves. They are plotted
in Fig.\ 6, where the displacement curve is also shown.
The inferred radius is 31$\pm$1 R$_\odot$. This is in very good
agreement with the prediction of recent period-radius
relation of Gieren et al. (1999) giving for P=3.29 days
a radius of 31.5 R$_\odot$. This relatively large radius
implies that CK~Cam
is a regular Type I Cepheid (i.e. young supergiant star), which
is also supported by the low galactic latitude (8.7$^\circ$).

The effect of phase-dependent velocity differences was found to
be much lower than the estimated systematic errors due to the
initial simplifications. By replacing V$_1$ to V$_2$ data
the resulting radius changed only 0.2 R$_\odot$ (0.6\%).

\section{Line profile variations}

Different line broadening mechanisms occuring in a pulsating
atmosphere can be globally described by the variation of the
Full Width at Half Minimum (FWHM). Recently, Gillet et al. (1999)
have used this parameter to trace the turbulent velocity
variations in $\delta$~Cep. From nonlinear, nonadiabatic
pulsational models they conclude that the main factor governing
the line broadening processes is the global compression/expansion
of the atmosphere. Shock wave effects turned out to be
much weaker in their models.

% Fig.7
\begin{figure*}
\begin{center}
\leavevmode
\psfig{figure=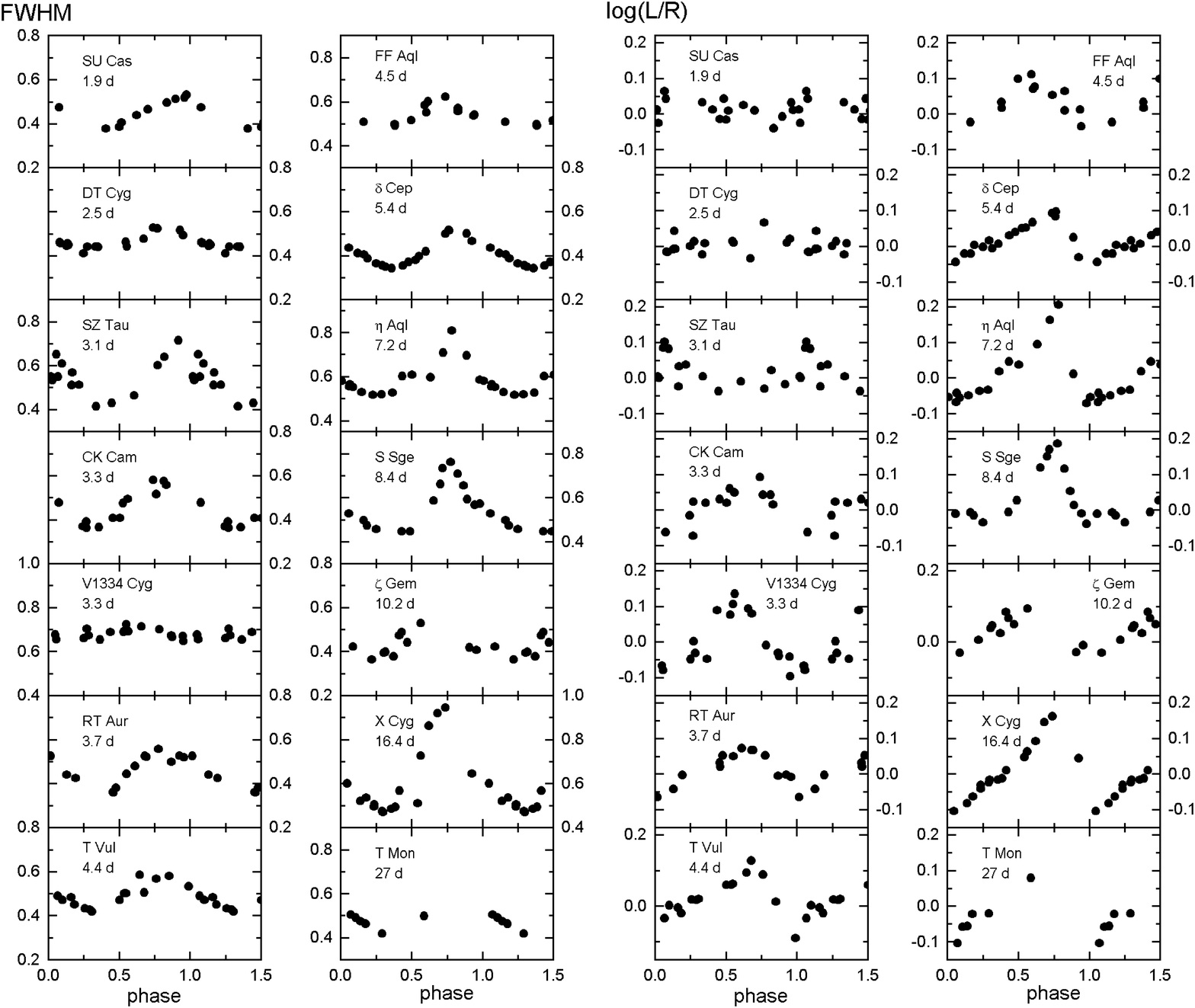,width=\textwidth}
\caption{FWHM and line profile asymmetry variations
of the iron line $\lambda$ 6411.658 \AA\ in 14 Cepheids.}
\end{center}
\end{figure*}

In order to get an overall picture about the period dependence
of this phenomenon, we have determined FWHM curves for 14 stars similar
to that of Gillet et al. (1999). We chose several unblended lines
of different excitation potentials (EP) and found the shape
of FWHM curves not depending on the actual EP value. We plotted
the resulting curves for 14 stars in Fig.\ 7 calculated for
line $\lambda$ 6411.658 \AA, which is a neutral iron line
of EP=3.65 eV.

The general trends are very similar for the majority of the stars.
The smallest FWHM always occurs very close to the phase of maximal radius.
On the other hand,
the largest FWHM is usually associated with the velocity reversal
point (around $\phi\approx$ 0.8--0.85). This is just before
the smallest radius, when the global compression is the
strongest, dominating the line broadening.

There are three deviating stars in our sample. Two of them (SU~Cas
and SZ~Tau) have the smallest line width in similar phases than in
other stars, however, the largest FWHM occurs later, between
$\phi$=0.95--1.00 (see Fig.\ 8).
These variables are thought to be first overtone pulsators (referred
as s-Cepheids)
which could be a likely explanation for the deviation.
Bersier \& Burki (1996) suggested a separation between
classical and s-Cepheids based on their radial velocity and
turbulence variations. This aspect of atmospheric
phenomena, i.e. the effects of overtone pulsation, has been
rarely studied theoretically. Evans et al. (1998) explained
the strong period variation of the first overtone Cepheid
Polaris by the complexities of the envelope acoustic cavity.
Nevertheless, there is a number of pieces of evidence for
a possible connection between the atmospheric turbulence phenomena
and the mode of pulsation. In our sample, there are two
other s-Cepheids (DT~Cyg and FF~Aql), which do not follow
the suggested distinction, therefore, we cannot draw a firm
conclusion on this issue.

The third deviating star is V1334~Cyg, which is a well-known binary
Cepheid with a bright companion. This star neither
follows the overall pattern of FWHM-variations (i.e. largest FWHM
occurs around $\phi$=0.8--0.9), nor fits the expectations on the
absolute value of the full-width. As can be seen in Fig.\ 7, its
line profile is quite stable around 0.67 \AA, while there is a slight
decrease around $\phi \approx$0.75.
A possible explanation for
these peculiar behaviour could be the effect of a bright
companion which contributes significantly to the observed
line profiles (see below).

The asymmetry of spectral lines of pulsating stars have been
extensively investigated by many authors, recently Sasselov
\& Lester (1990), Sabbey et al. (1995) and Albrow \& Cottrell (1994).
The asymmetry parameter (AP) was defined in several forms in these
papers. In the followings we adopt the form given by Sasselov \& Lester
(1990), namely the logarithm of left-to-right half-width ratio of
a given spectral line (in this case the Fe I $\lambda$ 6411.658 \AA).
We checked the dependence of AP on choosing the spectral line
and found no significant variation in the spectral region covered
by our observations. The precision of the AP values was tested
using some telluric lines which are expected to be close to
AP=0. The uncertainty of AP was found to be 1--2 percent.

The behaviour of AP during the pulsational cycle can be observed
in the right side of Fig.\ 7 for the programme stars. In the
case of SU~Cas, DT~Cyg and SZ~Tau only very weak variations can
be seen, while there are much larger changes in the case of
long-period Cepheids. The natural explanation is the smoother
pulsation in first overtone stars, where the amplitudes of
pulsation are smaller and the dynamic effects are weaker.
V1334~Cyg again deviates from other low-amplitude pulsators
showing asymmetry variation that is comparable to long-period
Cepheids. On the other hand, the large asymmetries occuring
in long-period stars are associated with large FWHM variations,
while there is no such phenomenon in V1334~Cyg.

% Fig.8
\begin{figure}
\begin{center}
\leavevmode
\psfig{figure=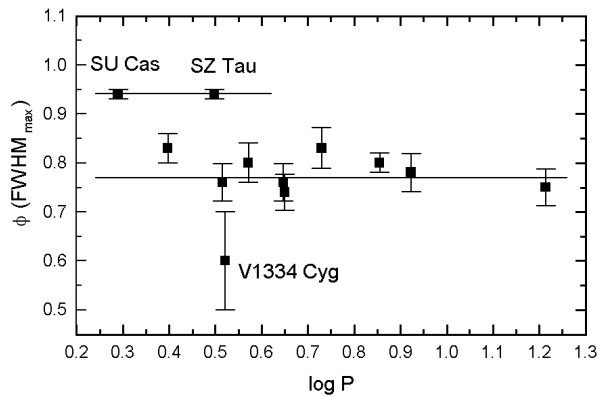,width=\linewidth}
\caption{The period dependence of the phase of maximal FWHM.
The well-separated positions of SU~Cas and SZ~Tau suggest a
possible effect of the mode of pulsation on the atmospheric
turbulence phenomena.}
\end{center}
\end{figure}

Generally, the AP variation curves resemble the radial velocity
curves indicating that the projection effect (the center of
the stellar disk shows maximal Doppler-shift, while the limb
has no Doppler-shift) contributes at least partly to the
observed asymmetries. On the other hand, the model
calculations by Sabbey et al. (1995) showed that there are
other factors that also cause line asymmetries (mainly
the varying depth of line formation over the pulsation cycle)
producing larger observed asymmetries. It is visible in Fig.\ 7, that
the amount of asymmetry variation (10--20 percent) is about the
same for all fundamental pulsators in our sample.
These new observations provide additional confirmation
of the theoretical results obtained by Sabbey et al. (1995) and
Albrow \& Cottrell (1994).

\subsection{The companion of V1334 Cygni}

% Fig.9
\begin{figure}
\begin{center}
\leavevmode
\psfig{figure=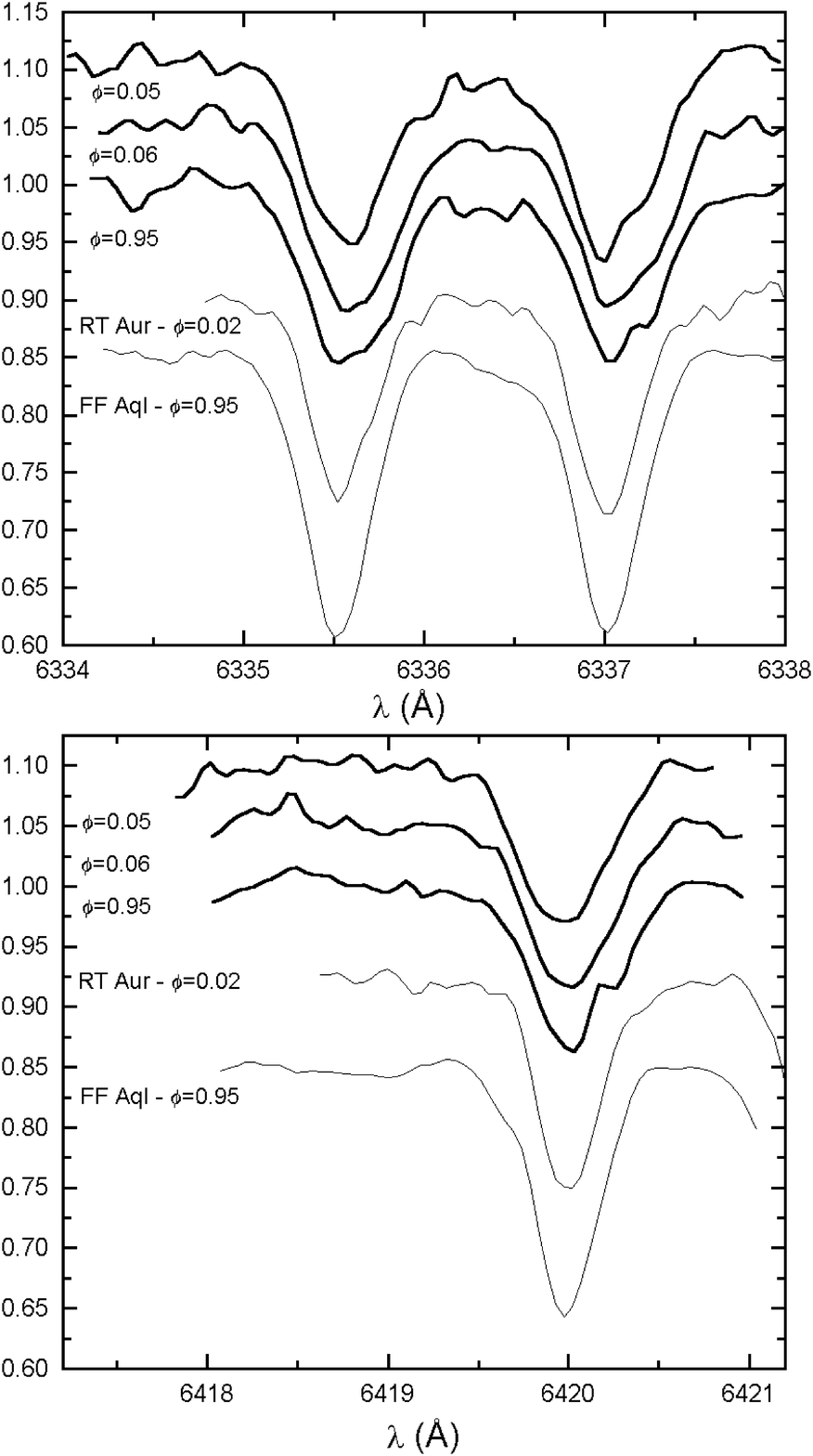,width=\linewidth}
\caption{Line doubling in V1334~Cyg compared with similar Cepheids in similar
pulsational phases.}
\end{center}
\end{figure}

The low-amplitude pulsator V1334~Cyg is unique among other
low-amplitude Cepheids in many respects, as was shown in
the previous sections. Here we try to construct a self-consistent
explanation of all these peculiarities by assuming a bright yellow
companion star moving in a close orbit around V1334~Cyg~A.
Both the unusually constant FWHM and large AP variation
can be explained with the presence of the spectral lines
of the component star with slightly Doppler-shifted to
the lines of V1334~Cyg~A. The secondary lines make the
combined line profiles broader, thus, the variation
of FWHM due to pulsation is obscured. Furthermore,
the Doppler-shift of the pulsation displace the Cepheid
lines from the companion lines producing higher asymmetries.
In parallel with these effects, the measured amplitude of the
radial velocity variation decreases because the secondary lines
move the centers of the combined lines closer to their
equilibrium positions. All these effects can be seen in a
numerical model
using simulated data that is a subject of a follow-up
publication (Kiss, in prep.).

We made a few very simple considerations on the possible nature
of the hypothetic component star. The visual amplitude of
V1334~Cyg is about 0.2 mag being too small compared to
other short-period Cepheids. If we consider V1334~Cyg to be
a fundamental mode pulsator, its amplitude should be
about 0.4--0.6 mag (see, e.g., CK~Cam, RT~Aur, T~Vul).
Assuming a virtual amplitude decrease due to the significant
amount of secondary light, one can compute the luminosity
ratio L(2)/L(Cepheid). In order to get an amplitude
of 0.2 mag instead of 0.5, a ratio of L(2)/L(Cepheid)$\approx$3
is required suggesting an unphysical yellow supergiant, which
is too bright and still undetected.
However, a more consistent result is given by the assumption
of first overtone pulsation. In that case the intrinsic
visual amplitude is about 0.4 mag (see, e.g., SU~Cas, DT~Cyg,
SZ~Tau, FF~Aql), and the observed 0.2 mag implies a ratio
L(2)/L(Cepheid)$\approx$1. This is still a very bright
component, but it gives more consistent picture of the system.
If it has a similar spectral type than V1334~Cyg~A does, than it
is almost invisible in the IUE spectra and consequently,
its detection is much more difficult in the ultraviolet.
The bright blue companion (B7.0V) cannot be responsible for
all these effects
because it is much fainter than the Cepheid itself in this spectral
region and also it is a fast rotator with v~sin~i around 200
km~s$^{-1}$ (Evans, personal communication).

We found some observations directly suggesting the presence of a
bright secondary component in the line profiles.
Fig.\ 9. shows line profiles of V1334~Cyg at certain phases
corresponding to the highest asymmetries (thick lines). As
a comparison, line profiles of other Cepheids with similar
period at the same phases are also plotted (thin lines).
It can be seen that a) the lines of V1334~Cyg are much
broader and b) a multiple structure is resolved in V1334~Cyg that
is absent in the other two Cepheids (note, that FF~Aql is also
a binary but its component is much fainter to be detected in this
spectral range). Both the broad lines and the multiple structure
strongly support the hypothesis of a bright yellow companion.

We did not observe systematic shift of the FWHM and AP curves between
1996 and 1997, while during this period the $\gamma$-velocity
of V1334~Cyg changed by 11 km~s$^{-1}$ due to orbital motion.
Consequently, the component that causes the orbital motion
cannot be the same that is responsible for the line profile
anomalies. We propose that there is a close, yellow component
of V1334~Cyg~A on an orbit with low inclination producing no
observable Doppler-shifts, but affecting the line profiles
of the pulsating star. These two stars move together along an
orbit around a massive third star with an amplitude of
27.8 km~s$^{-1}$ (Evans 1995). Because the blue component
detected by the IUE may not be that massive third star, the
system of V1334~Cyg might be a complicated multiple system.

\section{Summary}

The results presented in this paper can be summarized as follows.

1. We made high-resolution echelle spectroscopy for 18 northern Cepheids
in the yellow-red spectral region, between 5900 \AA\ and 6660 \AA.
New radial velocity data with internal accuracy of about 0.15 km~s$^{-1}$
were calculated using the digital cross-correlation technique.
Single-line radial velocities were also determined by the bisector
technique and their average values were compared with the ccf-data.
There is a good agreement at a level of 0.5--1 km~s$^{-1}$ between
these two techniques applied for the same observations.

2. We compared our new measurements with recently published spectroscopic
and CORAVEL-type velocities. We found {\it i)} perfect agreement with
conventional spectroscopic data and {\it ii)} systematic differences as
large as 1--3 km~s$^{-1}$ between CORAVEL and our data in certain phases.
Possible explanations for this discrepancy are the line profile perturbing
effects of velocity gradient and line asymmetries, although our
digital ccf-data are less affected than those of obtained by the CORAVEL
technique.

3. Line level effects were studied for a set of spectral lines with
different excitation potential. We found very clear examples of
line level effect for low-EP ($<$2 eV) and high-EP ($>$6 eV) lines.
The Si II 6347.095 \AA\ line is a very good indicator of the
strength of level effects. The observed level effect get stronger
with the increasing period, starting from 1--2 km~s$^{-1}$ for P$<$7 days
up to 6--8 km~s$^{-1}$ for P$>$7 days. Using these velocity differences
we estimated the velocity gradients and their dependence on the stellar
parameters. Our approximative approach gave a roughly constant
value of $\nabla v = 10^{-5} {\rm s}^{-1}$.

4. We performed a Baade-Wesselink analysis for CK~Cam discovered
by the Hipparcos satellite.
The inferred radius is 31$\pm$1 R$_\odot$ suggesting a regular
Type I Cepheid.

5. The phase dependent behaviour of FWMH and asymmetry variations
were examined. The fundamental and overtone pulsators seem to follow
different trends. In the case of
fundamental pulsation, the smallest FWHM always occurs very close
to the phase of maximal radius, while the largest FWHM is associated with
the global compression of the atmosphere. In the first overtone Cepheids
the largest FWHM is shifted toward the temperature maximum, around
$\phi=0.95-1.00$. The observed line asymmetries are consistent with
theoretical predictions involving the effects of projection and
varying depth of line formation over the pulsation cycle.

6. We present observational pieces of evidence for a bright,
yellow companion of V1334~Cyg based on the fine spectral peculiarities
of this Cepheid. This suspected decreases the measurable light and radial
velocity amplitudes by a factor of 2 and may explain the other
peculiarities of the line profile variations.

%\begin{acknowledgment}

\bigskip

\noindent This work has been supported by Hungarian E\"otv\"os Fellowship
to J.V., OTKA Grants T022259, F022249 and Foundation for Hungarian
Education and Science. The ADS Abstract Service was used to access references.
%\end{acknowledgment}

\end{document}